\documentclass[preprint, nofootinbib, tightenlines, eqsecnum, floats, aps,amsfonts]{revtex4}
\usepackage{epsfig,bbm}
\usepackage{amsmath}
\usepackage{amsfonts}
\usepackage{textcomp}

\def\mathbi#1{\textbf{\em #1}}

\begin{document}

\title{Ground-state wave-functional in (2+1)-dimensional Yang-Mills theory:\\Abelian limit, spectrum and robustness}

\author {Lionel Brits\footnote{E-mail:lionelbrits@phas.ubc.ca}}
\affiliation{Department of Physics and Astronomy, University of British Columbia, Vancouver, British Columbia, V6T 1Z1, Canada}

\begin{abstract}
We compute the glueball spectrum in (2+1)-dimensional Yang-Mills theory by analyzing correlators of the Leigh-Minic-Yelnikov ground-state wave-functional in the Abelian limit. The contribution of the WZW measure is treated by a controlled approximation and the resulting spectrum is shown to reduce to that obtained by Leigh et al., at large momentum.
\end{abstract}

\maketitle

\section{Introduction}

The study of non-Abelian gauge theories is complicated by the fact that by restricting oneself to states in the Hilbert space that are gauge invariant, or, alternatively, representatives of equivalence classes of states related by gauge transformations, one almost necessarily introduces interactions that complicate calculations. In perturbative treatments gauge invariance is achieved at the expense of the form of the bare propagator of the gauge-boson, or the introduction of ghosts, while in functional treatments  Gauss' law must either be imposed as a constraint, or solved implicitly through a change of variables, as is also the case in lattice calculations.

One promising solution to the constraint problem is the use of Wilson-line variables \cite{PhysRevD.10.2445}, for example in the work by Karabali and Nair on (2+1)-dimensional Yang-Mills theory \cite{1996NuPhB.464..135K}. The strength of their approach is that the field variables are encoded into a single variable along with a reality condition that allows it to be treated holomorphically, opening up the rich structure of complex analysis. Recent calculations by Leigh, Minic and Yelnikov based on this work have also yielded a candidate ground-state wave-functional as well as approximate analytical predictions for the \(J^{PC} = 0^{++}\) and \(J^{PC} =0^{--}\) glueball masses \cite{Leigh:2005dg} which are in good agreement with lattice calculations. However, the effect of the non-trivial configuration space measure, i.e., the WZW action, was omitted from calculations without sufficient justification.

In the following section we review the Karabali-Nair formalism and outline the procedure used by Leigh et al. to obtain the vacuum wave-functional and glueball spectrum. In part III we attempt a conservative approximation of the glueball spectrum that incorporates the WZW measure by expanding relevant operators about the Abelian limit. We then compare our results to that of \cite{Leigh:2005dg} and analyze the robustness of the solution.

\section{Background}

\subsection{Karabali-Nair formalism}

We consider \(SU(N)\) Yang-Mills in \((2+1)\)-dimensions in the Hamiltonian formalism, and denote the gauge potential by \(A_i = -i A_i^a t^a\), \(i = 1...2\), where the \(N\times N\) Hermitian matrices \(t^a\) generate the \(\mathfrak{su}(N)\) Lie algebra \([t^a,t^b] = if^{abc}t^c\) and we choose \( \operatorname{tr} (t^a t^b) = \frac{1}{2}\delta^{ab}\), while the gauge covariant derivative in the fundamental representation is \(D_i = \partial_i + A_i\). In the Hamiltonian formalism the component \(A_0\) is used up as a Lagrange multiplier in enforcing the Gauss' law constraint \(\mathbf{D}\cdot\mathbf{E} = 0\), and is subsequently ignored. Thus at this point we have yet to fix a gauge, so that the theory is invariant under (time independent) gauge transformations. We can write \(A_i\) in terms of path-ordered phases (Wilson lines) with one point at infinity, i.e.,
\begin{equation}
A_i = -(\partial_i M_{i} )M_{i}^{-1}, \mbox{ (no summation)},\label{Adefn}
\end{equation}
where
\begin{equation}
\label{Mdefn}
M_i(\mathbi{x}) = \mathcal{P} e^{-\int_{\infty}^\mathbi{x} dy^j A_j },
\end{equation}
and the integration contour is taken holding all but \(x^i\) fixed. Note that \(A_i\) is not a pure gauge since the matrices \(M_i\) are generally different. That the path-ordered phases encode at least as much information as the gauge potential is evident from equations (\ref{Adefn}, \ref{Mdefn}) but this can also be seen as follows: From knowledge of \(M_i\) everywhere we can construct infinitesimal holonomies which are proportional to \(F_{\mu\nu}\), from which we can reconstruct \(A_\mu\) in the coordinate gauge (\(x^\mu A_\mu = 0\)) \cite{Azam1991, Mendel1984}.

It is convenient to introduce the complex coordinates \(z = x - i y\) and \(\bar{z} = x + i y\). Correspondingly, \(\partial \equiv \frac{\partial}{\partial z} = \frac{1}{2}\left(\partial_1 + i \partial_2\right) \) and \(\bar{\partial} \equiv \frac{\partial}{\partial \bar{z}} =  \frac{1}{2}\left(\partial_1 + i \partial_2\right) \), while \(A \equiv A_z = \frac{1}{2}\left(A_1 + i A_2\right)\) and \(\bar{A} \equiv A_{\bar{z}} = \frac{1}{2}\left(A_1 - i A_2\right)\). We can again define matrices \(M_z\) and \(M_{\bar{z}}\) that satisfy \(A_z = -\partial_z M_z M_z^{-1}\) and \(A_{\bar z} = -\partial_{\bar z} M_{\bar z} M_{\bar z}^{-1}\) respectively. Noting that \(M_{\bar z} = M_z^{\dagger-1}\), we drop the subscript \(z\):
\begin{subequations}
\begin{equation}
\label{Azdefn}A = -\partial M M^{-1},
\end{equation}
\begin{equation}
\label{Azbdefn}\bar{A} = M^{\dagger-1}\bar\partial M^{\dagger}.
\end{equation}
\end{subequations}

The path-ordered phase \(M \in SL(N,\mathbb{C})\) transforms locally if we restrict ourselves to gauge transformations that are fixed (normally to \(1\)) at infinity. Under a time-independent gauge transformation \(A_i \to A_i^g = g A g^{-1} - \partial_i g g^{-1}\) where \(g \in SU(N)\), \(M\) transforms as \(M \to M^g = g M\). We are led to define a \textit{local} gauge-invariant variable \(H = M^\dagger M\). The change of variables from \((A,\bar A) \to H\) involves a Jacobian determinant so that the measure on the configuration space \(\mathcal{C}\) (the space of gauge potentials modulo allowable gauge transformations) is \cite{1996NuPhB.464..135K}
\begin{align}
\label{determinant}
d\mu(\mathcal{C}) &= \det (D \bar D) d\mu(H),\\
&= \left[ \frac{\det'(\partial \bar\partial )}{\int\!d^2x} \right]^{\dim G} e^{2\,c_A S_{WZW}[H]}d\mu(H),
\end{align}
where \(c_A\) is the quadratic Casimir in the adjoint representation, i.e., \( (T^c T^c)_{ab} \equiv -f^{cad} f^{cdb} = c_A \delta^{ab} \), equal to \(N\) for \(SU(N)\), and \(S_{WZW}[H]\) is the Wess-Zumino-Witten action,
\begin{align}
S_{WZW}[H] &= \frac{1}{2\pi}\int\!d^2x\,\operatorname{tr}(\partial H \bar\partial H^{-1})\nonumber\\
&+ \frac{i}{12\pi}\int\!d^3x\,\varepsilon^{\mu\nu\alpha}\operatorname{tr} (H^{-1}\partial_\mu H H^{-1}\partial_\nu H  H^{-1}\partial_\alpha H).
\end{align}
Careful regularization is involved in obtaining this result as the determinant has an anomaly. Inner products and expectation values are calculated using this measure, as
\begin{equation}
\label{Oexp}
\left\langle \Psi_1 | \mathcal{O} |\Psi_2 \right\rangle = \int\!d\mu(H) e^{2\,c_A S_{WZW}[H]} \Psi_1^*[H] \mathcal{O} \Psi_2[H],
\end{equation}
so that expectation values are essentially correlators of the Euclidean, Hermitian WZW theory. These, in turn, can be found (at least in principle) by solving the Knizhnik-Zamolodchikov equations for the unitary theory and performing an analytic continuation to the Hermitian case \cite{1996NuPhB.464..135K}. However, already at the four-point level one encounters non-trivial expressions involving hypergeometric functions \cite{Knizhnik:1984nr}, whereas the calculation wish to attempt contains an infinite series of such correlators (from expanding the wave-functional), with the next-leading-order being a six-point function. The need for approximation is therefore evident.

In the Hermitian WZW theory only correlators made up of integrable representations of the current algebra are well defined, so that all objects of interest can be written in terms of the WZW current \(J = \frac{c_A}{\pi} \partial H H^{-1}\) \cite{1998PhLB..434..103K}. This can also be seen from the observation that \((-\partial H H^{-1}, 0)\) is a field-dependent \(SL(N,\mathbb{C})\)-valued gauge transformation of \((A, \bar{A})\). Since the Gauss' law operator generates gauge transformations and vanishes on physical states, we may everywhere let \((A, \bar{A}) \to (-\partial H H^{-1}, 0) \). In particular, the Hamiltonian \(H = \int\!d^2x\,\operatorname{tr}\left( g^2 E_i^2 + \frac{1}{g^2} B^2\right)\) in terms of the variable \(J\) can be written as \cite{1998PhLB..434..103K}
\begin{equation}
\label{HKKN}
H = m\left[  \int_x\! J^a(\mathbi{x}) \frac{\delta}{\delta J^a(\mathbi{x})}  + \int_x\!\int_y\! \Omega^{ab}(\mathbi{x}-\mathbi{y}) \frac{\delta}{\delta J^a(\mathbi{x})} \frac{\delta}{\delta J^b(\mathbf{y})}   \right] + \frac{\pi }{m c_A}\int_x\!\bar{\partial}J^a(\mathbi{x}) \bar{\partial}J^a(\mathbi{x}),
\end{equation}
where \(m \equiv \frac{g^2 c_A }{2 \pi}\), \( \Omega^{ab}(\mathbi{x}-\mathbi{y}) \equiv \frac{c_A}{\pi}D^{ab}_x\bar{G}(\mathbi{y}-\mathbi{x}) \) and the Green's function \(\bar G\) is defined through \( \bar\partial_y \bar G(\mathbi{y}-\mathbi{x}) = \delta^{(2)}(\mathbi{y}-\mathbi{x})\).

For physical states, \(\Psi_{phys}[A,\bar A] = \Psi_{phys}\left[-\frac{\pi}{c_A}J,0\right]\), so that wave-functionals constructed out of \(J\) are automatically gauge invariant. The price paid for implicitly solving the constraint \(\mathbf{D}\cdot\mathbf{E} = 0\) is a local, ``holomorphic'' invariance of equations (\ref{Azdefn},\ref{Azbdefn}) under \(M(z,\bar z) \to M(z,\bar z) h(z), M^\dagger(z,\bar z) \to h^\dagger(\bar z)M(z,\bar z) \) where \(h(z)\) is a unitary matrix depending only on \(z\). Physical states must therefore also be holomorphically invariant.

\subsection{Vacuum Wave-functional}

Leigh et al., proposed the following \textit{Ansatz} for the vacuum wave-functional \cite{Leigh:2005dg}:
\begin{equation}
\label{LeighWaveFunctional}
\Psi[J] = \exp\left(-\frac{\pi}{2 c_A m^2} \int\! \bar\partial J  K(L) \bar\partial J \right) + ...,
\end{equation}
where the kernel \(K\) is a power series in the holomorphic-covariant derivative \( \Delta \equiv \frac{ \{D, \bar \partial \} }{2} \equiv m^2 L\). Terms in the exponent of higher order in \(\bar\partial J\) that can't be absorbed into the kernel are denoted by ellipsis; therefore, the \textit{Ansatz} is not the most general one. Note that \( \bar\partial J  K(L) \bar\partial J = \bar\partial J  K\left( \frac{\bar\partial \partial}{m^2} \right) \bar\partial J + O(J^3)\). Writing \(\Psi = e^P\), the action of kinetic energy term in (\ref{HKKN}) on \(\Psi\) becomes
\begin{equation}
\label{TP}
T \Psi = \left[T P + m \int_x\!\int_y\! \Omega^{ab}(\mathbi{x}-\mathbi{y}) \frac{\delta P}{\delta J^a(\mathbi{x})} \frac{\delta P}{\delta J^b(\mathbf{y})}  \right] \Psi. 
\end{equation}
To second order in \(\bar\partial J\), the second term in brackets yields
\begin{equation}
\label{Tsecond}
\frac{\pi}{c_A m}\int\!d^2x\, \bar\partial J \left[\frac{\partial\bar\partial}{m^2} K^2\left( \frac{\partial\bar\partial}{m^2}\right)\right] \bar\partial J + ...
\end{equation}
The action of \(T\) on terms of the form \(\mathcal{O}_n \equiv \int\bar\partial J (\Delta^n) \bar\partial J \) is less straight-forward. In reference \cite{2006hep.th....4060L} Leigh et al. argue that holomorphic invariance requires mixing between terms of different order in \(\bar\partial J\) in such a way that \(T \mathcal{O}_n = (2+n)m\mathcal{O}_n + ...\), with the result explicitly demonstrated for \(\mathcal{O}_0\) and \(\mathcal{O}_1\). Whether this result acquires corrections for higher values of \(n\) is not known. Formally, then, the  Schr\"odinger becomes
\begin{equation}
\label{Tformal}
H \Psi  = \left[\frac{\pi}{c_A m} \int \bar\partial J \left( - \frac{1}{2L} \frac{d}{dL} \left[L^2 K(L)\right] + L K^2(L) + 1\right) \bar\partial J  \right]\Psi = E \Psi.
\end{equation}
The eigenvalue equation
\begin{equation}
- \frac{1}{2L} \frac{d}{dL} \left[L^2 K(L)\right] + L K^2(L) + 1 = 0
\end{equation}
is then solvable using the substitution \(K = -\frac{p'}{2p}\) which casts it into Bessel form. The unique, normalizable solution with correct UV asymptotics was found by Leigh et al., to be \cite{Leigh:2005dg}:
\begin{equation}
K(L) = \frac{1}{\sqrt{L}} \frac{J_2(4\sqrt{L}) } {J_1(4\sqrt{L}) }.
\end{equation}
\subsection{Mass Spectrum}
In order to extract meaningful information from the ground state \(\Psi\) it is necessary to compute correlation functions as in equation (\ref{Oexp}). In \cite{2006hep.th....4060L} the operator \(\operatorname{tr}\bar\partial J \bar\partial J\) was found to be even under parity and charge conjugation, and so creates \(0^{++}\) states. A good starting point then is the calculation of \(\left\langle (\bar \partial J \bar \partial J )_x (\bar \partial J \bar \partial J )_y \right\rangle\). As we have already noted, such a computation is presently intractable. However, in \cite{2006hep.th....4060L, Leigh:2005dg} it is argued that in the large-\(N\) limit, the variables \(\bar\partial J\) represent the correct physical degrees of freedom so that integration over these variables can be done as in a free theory, in the sense that
\begin{equation}
\left\langle \bar \partial J^a_x \bar \partial J^b_y \right\rangle \sim \delta^{ab} K^{-1}(|\mathbi{x}-\mathbi{y}|).
\end{equation}
\(K^{-1}(|\mathbi{x}-\mathbi{y}|)^2\) is then identified with a particular two-point function probing \(0^{++}\) glueball states, so that the mass spectrum of the vacuum can be read off from the analytic structure of \(K^{-1}(k)^2\). Essentially, it is argued that in the \(\bar\partial J\) configuration space the WZW measure can be neglected. The interpretation of this statement will be explored further in this paper. 

The asymptotic form of \(K^{-1}(|\mathbi{x}-\mathbi{y}|)^2\) was found to be \cite{Leigh:2005dg}
\begin{equation}
K^{-1}(|\mathbi{x}-\mathbi{y}|)^2  \to \frac{1}{32 \pi |\mathbi{x}-\mathbi{y}|} \sum_{n,\, m =1}^{\infty} (M_n M_m)^{3/2}e^{-(M_n+M_m)|\mathbi{x}-\mathbi{y}|},
\end{equation}
where \(M_n \equiv \frac{j_{2,n} m}{2}, n = 1,2,3...\) and \(j_{2,n}\) are the zeros of \(J_2(z)\). Comparing to the two-point function of the free Boson, equation (\ref{freebosonpropagator}), we can see that the \(0^{++}\) states have masses given by various combinations of \(M_n\). This result is in good agreement with lattice calculations presented in \cite{teper-1999-59, lucini-2002-66} and in the next section we will attempt to give justification for this agreement in light of the approximations that have been made.

\section{Abelian Expansion}
The statement that integration over the variables \(\bar\partial J\) can be done explicitly should not be interpreted literally, as the change of variables from \(H\) to \(\bar\partial J\) involves a further factor \(\det (D\bar \partial)^{-1}\) where \(D\) is now the holomorphic covariant derivative with \(J\) as connection. Although the derivatives \(D\) and \(\bar\partial\) are related to the original expressed in terms of \((A,\bar A)\) through conjugation by \(M^{\dagger-1}\), the determinant also suffers from a multiplicative anomaly which is expressed by the Polyakov-Wiegmann identity that relates \(S_{WZW}[gh]\) to \(S_{WZW}[g]\) and \(S_{WZW}[h]\) (see \cite{Polyakov:1983tt, Polyakov:1984et}). Therefore we shall approach the problem by exploring it in a particular limit where the WZW action at least allows for tractable calculations. 

We now attempt to calculate \( \left\langle (\bar \partial J \bar \partial J )_x (\bar \partial J \bar \partial J )_y \right\rangle\) by performing the path integral in equation (\ref{Oexp}) in the Abelian limit. Following \cite{1998NuPhB.524..661K} we write \(H = e^{\varphi}\) and expand terms of the form \(H^{-1}f(\varphi)H\) in powers of \(\varphi\). In the adjoint representation, \(\varphi^{ab} = f^{abc} \varphi^c\), so that an expansion in \(\varphi\) is necessarily an expansion in the structure constants. In the limit of small \(\varphi\) (see e.g., \cite{1998NuPhB.524..661K, 2004hep.th....7051L}), we can write the wave-functional and WZW factor in Gaussian form and calculate the four-point function as in a Euclidean field theory with action given by \(2c_A S[H] + \ln \Psi^*[H] \Psi[H]\). In all cases we expand terms inside the exponential, but not the exponential itself, i.e., we are performing a selective resummation \cite{1998NuPhB.524..661K}.  Then \(\left\langle (\bar \partial J \bar \partial J )_x (\bar \partial J \bar \partial J )_y \right\rangle\) gives, up to a disconnected diagram, the square of the two-point function:
\begin{equation}
\left\langle (\bar \partial J \bar \partial J )_x (\bar \partial J \bar \partial J )_y \right\rangle = 2 \left\langle \bar \partial J_x \bar\partial J_y \right\rangle \left\langle \bar \partial J_x \bar\partial J_y \right\rangle.
\end{equation}
The measure factor becomes the exponential of the free complex scalar field action, \(e^{2 c_A S[H]} \approx e^{-\frac{c_A}{2\pi} \int\!d^2z\,\partial\varphi^a \bar\partial\varphi^a}\). The complete measure factor can be integrated, so that the volume of the configuration space \(\mathcal{C}\) is finite, which leads to the existence of a mass-gap \cite{1998NuPhB.524..661K}. By approximating the WZW factor in this way we hope to make the calculation tractable and at the same time capture non-trivial effects. Already we notice that in the Abelian limit the zero mode causes the volume of \(\mathcal{C}\) to diverge - it was already noted in \cite{1998NuPhB.524..661K} that the WZW factor cannot be obtained in the Abelian limit; therefore the approximation may break down at low momentum.

To expand the wave-functional, note that \cite{Georgi1999}
\begin{align}
J &= \frac{c_A}{\pi} \int_0^1\!ds\, e^{s \varphi^a t^a} (\partial\varphi^b t^b) e^{-s \varphi^c t^c},\label{Jabelian1}\\
 &\approx \frac{c_A}{\pi} \partial\varphi + ...\label{Jabelian2},
\end{align}
where we have dropped higher powers of \(\varphi\), so that \(\bar\partial J \approx \frac{c_A}{\pi}\frac{\nabla^2}{4}\varphi\). That we are truly in the Abelian limit can be seen by allowing terms inside equation (\ref{Jabelian1}) to commute, thereby obtaining equation (\ref{Jabelian2}) exactly. Also, \((L \varphi)(\mathbf{k})\approx -\frac{\mathbf{k}^2}{4m^2} \varphi(\mathbf{k})\), so that the exponent of the wave-functional becomes
\begin{equation}
\int\!d^2z\, \bar\partial J^a K\left(\frac{\partial\bar\partial}{m^2}\right) \bar\partial J^a\, \approx \left(\frac{c_a}{\pi}\right)^2\int\!d^2k\, \varphi^a(\mathbf{k})\frac{\mathbf{k}^2}{4} K\left(-\frac{\mathbf{k}^2}{4m^2}\right) \frac{\mathbf{k}^2}{4} \varphi^a(-\mathbf{k}).
\end{equation}
Finally,
\begin{equation}
e^{2 c_A S[H]}\Psi^*[H] \Psi[H] \approx \exp\left\{-\frac{c_A}{2\pi}\int\!d^2k\, \varphi^a(\mathbf{k})\frac{2\mathbf{k}^2}{4}\left[\frac{2}{\mathbf{k}^2} + \frac{1}{m^2} K\left(-\frac{\mathbf{k}^2}{4m^2}\right) \right] \frac{\mathbf{k}^2}{4} \varphi^a(-\mathbf{k}) \right\}.
\end{equation}
To calculate \(\left\langle \bar \partial J(x) \bar \partial J (y)\right\rangle\), it is sufficient to know \(\left\langle \varphi^a (x) \varphi^b (y) \right\rangle\), as the former can be obtained from the latter by repeated differentiation. Therefore we have to analyze the field theory with effective kernel given by
\begin{equation}
\mathcal{K}\left(-\frac{\mathbf{k}^2}{4m^2}\right) \equiv \frac{1}{2} \frac{4 m^2}{\mathbf{k}^2} + K\left(-\frac{\mathbf{k}^2}{4m^2}\right).
\end{equation}

\subsection{Analytic Structure}
Let us begin by writing the effective kernel in terms of the formal parameter \(y \equiv 4\sqrt{L}\):
\begin{align}
\mathcal{K}(L) &= -\frac{1}{2}\frac{1}{L} + \frac{1}{\sqrt{L}}\frac{J_2(4\sqrt{L})}{J_1(4\sqrt{L})}\\
&= 	\frac{4}{ y}\left[-\frac{2}{y} + \frac{J_2(y)}{J_1(y)} \right].
\end{align}
As a first step towards finding the inverse of the kernel, consider the identity \cite{Milton1965}
\begin{equation}
\label{Besselsum}
J_{\nu-1}(y) + J_{\nu+1}(y) = \frac{2\nu}{y}J_{\nu}(y),
\end{equation}
from which we immediately see that
\begin{equation}
\mathcal{K}(L) = -\frac{4}{y}\frac{J_0(y)}{J_1(y)}.
\end{equation}
Another identity of interest is 
\begin{equation}
\frac{J_1(y)}{J_0(y)}= 2 y\sum_{s=1}^{\infty} \frac{1}{ j^2_{0,s}-y^2 },
\end{equation}
where \(j_{\nu,s}\) denotes the \(s^{th}\) zero of \(J_\nu(y)\). This result can be derived using common Bessel function identities along with the infinite product representation,
\begin{equation}
\label{Jinfprod}
J_\nu(y) = \frac{\left(\frac{1}{2}y\right)^\nu} { \Gamma(\nu+1)} \prod_{s=1}^\infty\left(1-\frac{y^2}{j^2_{\nu,s}}\right),
\end{equation}
and the fact that \(J_\nu(y)\) is a meromorphic function. Then
\begin{align}
\mathcal{K}(L)^{-1} = \frac{1}{2 } \sum_{s=1}^{\infty} \frac{ y^2}{y^2 - j^2_{0,s}}.
\end{align}
We therefore take the inverse effective kernel to have the following analytical structure
\begin{align}
\mathcal{K}^{-1}\left(-\frac{\mathbf{k}^2}{4m^2}\right) &= \frac{1}{2} \sum_{s=1}^{\infty} \frac{y^2}{y^2 - j^2_{0,s}},\\
&= \frac{1}{2} \sum_{s=1}^{\infty} \left(1 -   \frac{M_s^2}{\mathbf{k}^2 + M_s^2 }\right),
\end{align}
where \(M_s \equiv \frac{j_{0,s} m}{2}\). Following the treatment of \cite{Leigh:2005dg}, the real space  kernel has the following asymptotic behaviour:
\begin{align}
\mathcal{K}^{-1}(|\mathbi{x}-\mathbi{y}|) &= -\frac{1}{2} \sum_{s=1}^{\infty} \frac{   M_s^2}{2\pi} K_0\left(M_s |\mathbi{x}-\mathbi{y}|\right),\\
&\to -\frac{1}{4} \sum_{s=1}^{\infty}   M_s^{\frac{3}{2}} \frac{1}{\sqrt{2\pi |\mathbi{x}-\mathbi{y}|}} e^{-M_s |\mathbi{x}-\mathbi{y}|},
\end{align}
while the four-point function is
\begin{equation}
\mathcal{K}^{-1}(|\mathbi{x}-\mathbi{y}|)^2 = \frac{1}{32\pi |\mathbi{x}-\mathbi{y}|} \sum_{r,s=1}^{\infty} (M_r M_s)^{\frac{3}{2}}   e^{-(M_r+M_s) |\mathbi{x}-\mathbi{y}|}.
\label{Kminus2}
\end{equation}
Comparing to the propagator of the free Boson evaluated at fixed time, equation (\ref{freebosonpropagator}), we see that a multitude of particles with masses \(M_r + M_s\) have been identified. A comparison between our masses and those obtained in \cite{Leigh:2005dg} is given in table \ref{Comparemasstable}. Large-N lattice results \cite{teper-1999-59, lucini-2002-66, 2003NuPhB.668..111M} given for comparison in \cite{Leigh:2005dg} have also been reproduced. We have omitted the spurious pole due to \(j_{0,1}\) as it has no obvious value to compare to. As noted in \cite{Leigh:2005dg}, the discrepancy with the \(0^{++*}\) lattice result may suggest that this is in fact two states (corresponding to \(M_1 + M_2\) and \(M_2 + M_2\)) which are not resolved by the lattice calculation, or it may indicate a low-momentum breakdown of our calculation.

\begin{table}
\caption{Comparison of 0++ glueball masses given in units of string tension $\sqrt{\sigma} \approx\sqrt{\frac{\pi}{2}} $}
\label{Comparemasstable}
\begin{ruledtabular}
\begin{tabular}{llll}
State & Lattice\footnote{See \cite{teper-1999-59, lucini-2002-66,2003NuPhB.668..111M}} & Leigh, et., al. & Our rediction\\
\hline
\(0^{++}\) & \(4.065 \pm 0.055\) & 4.10 &4.40 \\
\(0^{++*}\) & \(6.18 \pm 0.13\) & 5.41 &5.65 \\
\(0^{++*}\) & & 6.72 &6.90 \\
\(0^{++**}\) & \(7.99 \pm 0.22\) & 7.99 &8.15\\
\(0^{++***}\) & \(9.44 \pm 0.38\) & 9.27 &9.40\\
\end{tabular}
\end{ruledtabular}
\end{table}

\section{Discussion}
The calculation presented here shows that in the Abelian limit the mass spectrum probed by \(J^{PC} = 0^{++}\) operators comprises sums of pairs of zeros of \(J_0(y)\), in contrast with the result of Leigh et al., which expresses these masses in terms of zeros of \(J_2(y)\). Thus we find a correction due to the inclusion of the WZW action to lowest order in \(f^{abc}\). Asymptotically \(J_\nu(y) \rightarrow \sqrt{\frac{2}{\pi y}} \cos \left( y-\frac{\nu\pi}{2} - \frac{\pi}{4} \right)\) so that \(J_2(y) \to -J_0(y)\) as \(y \to \infty\). Therefore our results reproduce those in \cite{Leigh:2005dg} at large momentum and give justification for approximations made therein. That the results agree at large momentum suggests that the analytic structure of the wave-functional is rather robust against short-distance corrections. At low momentum (small \(y\)), however, an additional pole arises due to \(j_{0,1}\), which is not seen in \cite{Leigh:2005dg}, and gives a constituent mass of \(M_1 \approx 0.96\sqrt{\sigma}\). Combinations using \(M_1\) do not seem to appear in lattice calculations presented in \cite{teper-1999-59, lucini-2002-66}. Whether this signals a breakdown of the Abelian approximation at low momentum that can be corrected by continuing the expansion to higher order is a possible direction for future investigation. 

\begin{acknowledgments}
The author would like to thank Moshe Rozali for guidance, as well as insight and many useful discussions. This work was supported by the National Science and Engineering Research Council of Canada.

\end{acknowledgments}

\appendix
\section{Free Boson in (2+1)-Dimensions}
In order to make a connection between the correlators obtained in the Hamiltonian formalism, e.g., equation (\ref{Kminus2}), with the (covariant) K\"allen-Lehmann spectral representation, we need to recognize the analytic structure of a 1-particle state when expressed non-covariantly. To do this, we analyze the two-point function of the free Boson for purely spatial separation:
\begin{align}
\label{freebosonpropagator}
\Delta_F(\mathbi{x}-\mathbi{y}) &= \int\!\frac{d^2k}{(2\pi)^2} \frac{1}{\sqrt{k^2 + m^2}} e^{i \mathbf{k}\cdot\mathbf{r}}\nonumber\\
&= \int_0^\infty \!\frac{k\,dk }{(2\pi)^2} \frac{2\pi J_0(k r)}{\sqrt{k^2 + m^2}}\nonumber\\
&= \frac{1}{2\pi |\mathbi{x}-\mathbi{y}|}e^{-m|\mathbi{x}-\mathbi{y}| }
\end{align}

\bibliography{abelian}        

\end{document}